\documentclass[epj]{webofc}%
\usepackage[utf8]{inputenc}
\usepackage[varg]{txfonts}
\usepackage{booktabs}
\usepackage{xcolor}
\usepackage[bookmarks,linktocpage,colorlinks,     linkcolor = darkred,
urlcolor  = darkblue,     citecolor = darkgreen]{hyperref}
\usepackage{subfigure}
\usepackage{amsmath}
\usepackage{amsfonts}
\usepackage{amssymb}
\ifx\msidvipdf\msiundefined
\usepackage{graphicx}
\else%
\usepackage[dvipdfm]{graphicx}
\fi
\providecommand{\U}[1]{\protect\rule{.1in}{.1in}}
\definecolor{darkred}{rgb}{0.4,0.0,0.0}
\definecolor{darkgreen}{rgb}{0.0,0.4,0.0}
\definecolor{darkblue}{rgb}{0.0,0.0,0.4}
\wocname{EPJ Web of Conferences}
\woctitle{Lattice2017}

\ifx\pdfoutput\relax\let\pdfoutput=\undefined\fi
\newcount\msipdfoutput
\ifx\pdfoutput\undefined\else
\ifcase\pdfoutput\else
\ifx\paperwidth\undefined\else
\ifdim\paperheight=0pt\relax\else\pdfpageheight\paperheight\fi
\ifdim\paperwidth=0pt\relax\else\pdfpagewidth\paperwidth\fi
\fi\fi\fi
\ifx\msidvipdf\msiundefined\else
\AtBeginDvi{}
\fi
\begin{document}

\selectlanguage{english}

\begin{flushright}
\begin{minipage}{40mm}
\normalsize
KEK Preprint 2017-37 \\
CHIBA -EP-227 
\end{minipage}
\end{flushright}

\title{Lattice study of area law for double-winding Wilson loops}
\author{\firstname{Akihiro} \lastname{Shibata}\inst{1}\fnsep
\thanks{Speaker, \email{akihiro.shibata@kek.jp} } \and
\firstname{Seikou} \lastname{Kato}\inst{2}\and
\firstname{Kei-Ichi} \lastname{Kondo}\inst{3} \and
\firstname{Ryutaro} \lastname{Matsudo}\inst{4} }

\institute{
Computing Research Center, High Energy Acceleration Research Organization (KEK), Oho 1-1, Tsukuba 305-0801, Japan \and
Oyama National College of Technology, Oyama 323-0806, Japan \and
Department of Physics, Faculty of Science, Chiba University, Chiba 263-8522, Japan \and
Department of Physics, Faculty of Science and Engineering, Chiba University, Chiba 263-8522, Japan }%
\abstract{%
We study the double-winding Wilson loops in the SU(N) Yang-Mills theory on the lattice. We discuss how the area law falloff of the double-winding Wilson loop average is modified by changing the enclosing contours C1 and C2 for various values of the number of color N. By using the strong coupling expansion, we evaluate the double-winding Wilson loop average in the lattice SU(N) Yang-Mills theory. Moreover, we compute the double-winding Wilson loop average by lattice Monte Carlo simulations for SU(2) and SU(3). We further discuss the results from the viewpoint of the Non-Abelian Stokes theorem in the higher representations.
}
\maketitle

\section{Introduction}

\label{sec:intro} The Wilson loop is a gauge-invariant and important operator
for the lattice study. By using a single winding Wilson loop, we investigate
the static potential and the flux tube between quark and antiquark in the
fundamental representation. We further investigate the evidence of the dual
superconductivity such as the restricted field dominance and the magnetic
monopole dominance for the string tension, and dual Meissner effect.

There still exist two promising mechanisms for quark confinement. One is the
dual superconductivity \cite{DualSuper} in which the magnetic monopole plays a
dominant role for confinement. The other is the vortex picture in which the
center vortex plays a relevant role for confinement \cite{Greensite11}.
Recently, Greensite and H\"{o}lwieser presented the testing method for the
mechanism of confinement by using the double-winding Wilson loop which
enclosing contours are in the same plain \cite{Greensite15}. For the $SU(2)$
case, they investigate the average of the double-winding Wilson loop made of
Yang-Mills field, the center field extracted in the maximal center gauge, and
the Abelian-projection field in the maximal Abelian gauge. They showed that
the string tension for the minimum surface of the Wilson loop is the
difference of area behavior in case of the center-projection field as well as
the Yang-Mills field and the center-projection. In case of Abelian-projection
field, the string tension is the sum of area behavior in the same way as the
Abelian case. However, it must be examined whether replacing the Yang-Mills
field with the Abelian-projected field in the Wilson loop operator leads to
the correct result or not in view of the non-Abelian Stokes theorem.

In this talk, we investigate the double-winding Wilson loops of SU(N)
Yang-Mills theory on the lattice to know the correct behavior of the
expectation values such as the gauge group dependence, the relation to
N-ality, and the relation to the non-Abelian stokes theorem in view of  the
dual superconductivity.

\section{Double-winding \ Wilson loop}

\label{sec:double-winding}
\begin{figure}[bt] \centering
\includegraphics[origin=center,width=55mm]{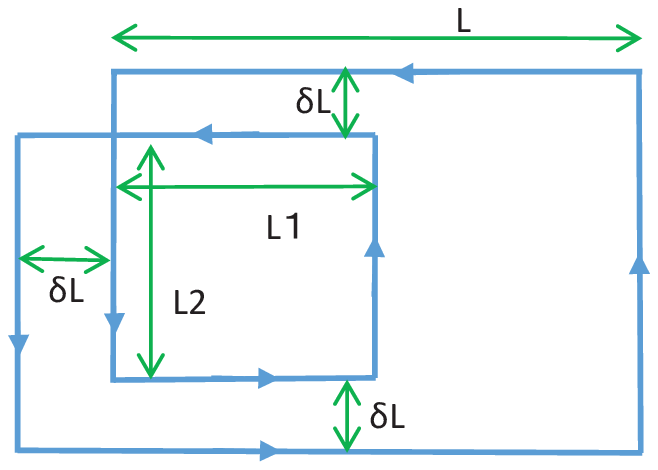}%
\includegraphics[origin=center, clip,width=40mm,]{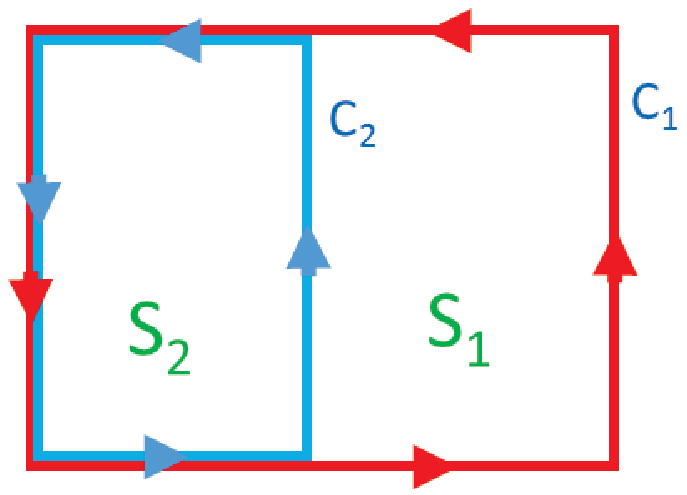}%
\includegraphics[origin=center, clip,width=40mm,]{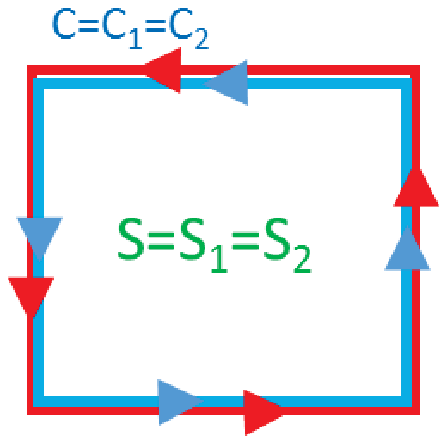}%
\caption{The contour of the double-winding Wilson loop.
The contour $C= C_1 \times
C_2$  winds once around a loop $C_1$ and once around a loop $C_2$
in the same direction.  The loop $C_1$ lies entirely in the minimal area of the loop $C_2$.
$S_1$ and $S_2$ represent the minimum areas formed by $C_1$ and $C_2$, respectively.
(leftmost) The parametrization of the contours of the double-winding Wilson loop.
(center) The $\delta L=0$ case.
(rightmost)  The the double-winding Wilson loop with the identical countour, i.e.,  the two identical loop, , $C_1=C_2$ }%
\label{fig:dwinding-wloop-L}%
\end{figure}

We set up the double-winding Wilson loop on the
lattice $W(C)$ (see Figure \ref{fig:dwinding-wloop-L}). The contour $C=$
$C_{1}\times C_{2}$ winds once around a loop $C_{1}$ and once around a loop
$C_{2}$ in the same direction, where the two coplanar loops $C_{1}$ and
$C_{2}$ share one point in common. The loop $C_{2}$ lies entirely in the
minimal area of the loop $C_{1}$. The area $S_{1}$ represents the minimum area
formed by $C_{1}$, i.e., $S_{1}=L\times L_{2}+\delta L\times(2L+L_{2}+\delta
L)$. The area $S_{2}$ represents the minimum areas formed by $C_{2}$, i.e.,
$S_{2}=L_{1}\times L_{2.}$ \ For simplicity of the analysis, we sometimes use
the case of $\delta L=0\,$(the center plane). The rightmost panel represents
the double-winding Wilson loop with an identical contour (two identical loops,
$C_{1}=C_{2}$, or $\delta L=0$ and $L_{1}=L$).

We investigate how the area law falloff of the double-winding Wilson loop
average is modified by changing the enclosing contours $C_{1}$ and $C_{2}$ for
various values of the number of color $N$ in $SU(N)$-Yang-Mills theory. We
first study the double-winding Wilson loop average by using the string
coupling expansion. Then, we evaluate the double-winding Wilson loop average
for $SU(2)\ $and $SU(3)$ by using Monte-Carlo simulations to examine the
result of the strong-coupling expansion and \ the calculation in the continuum
theory \cite{MastsudoKondo17}.

\section{Strong Coupling expansion}

\label{sec:SCexpansion} By using the strong coupling expansion \cite{SCexp},
we evaluate the double-winding Wilson loop average in the lattice $SU(N)$
Yang-Mills theory. For simplicity of calculation, we consider the $\delta L=0$
case of Fig. \ref{fig:dwinding-wloop-L}. We adopt the Wilson standard action
\begin{equation}
S_{g}=\beta\sum\limits_{x,\mu}\operatorname{Re}\mathrm{tr}\left(
1-U_{p}\right)  \text{, \ }U_{p}=U_{x,\mu}U_{x+\mu,\nu}U_{x+\nu.\mu}^{\dag
}U_{x,\nu}^{\dag}%
\end{equation}
where $U_{x,\mu}$ ($\in SU(N)$) is a gauge link variable, and $\beta=2N/g^{2}$
is the gauge coupling parameter. We consider the case $\beta\ll1$ ($g\gg1$)
and the average of the double-winding Wilson loop average is calculated by
expansion of $\beta.$ For simplicity of calculation, we investigate the case
of $\delta L=0$, i.e., the center panel of Fig.\ref{fig:dwinding-wloop-L}. The
$SU(N)$ group integrals are given as follows \cite{M.Crreutz1983}%
\cite{M.Cretz1978}
\begin{subequations}
\begin{align}
&  \int dU=1\\
&  \int dUU_{ab}=0\\
&  \int dUU_{ab}U_{cd}^{\dag}=\frac{1}{N}\delta_{ad}\delta_{bc}\\
&  \int dUU_{a_{1}b_{1}}U_{a_{2}b_{2}}\cdots U_{a_{N}b_{N}}=\frac{1}%
{N!}\epsilon_{a_{1}a_{2}\cdots a_{N}}\epsilon_{b_{1}b_{2}\cdots b_{N}}\\
&  \int dUU_{a_{1}b_{1}}U_{a_{2}b_{2}}\cdots U_{a_{M}b_{M}}=0\text{
\ \ (}M\neq0\text{ mod }N\text{ )}\\
&  \int dUU_{ab}U_{cd}U_{ij}^{\dag}U_{kl}^{\dag}=\frac{1}{4}\delta
_{2,N}\epsilon_{ac}\epsilon_{bd}\epsilon_{ik}\epsilon_{jl}\nonumber\\
&  \qquad+\frac{1}{N^{2}-1}\left[  \delta_{aj}\delta_{bi}\delta_{cl}%
\delta_{dk}+\delta_{al}\delta_{bk}\delta_{ci}\delta_{di}-\frac{1}{N}\left(
\delta_{aj}\delta_{bk}\delta_{cl}\delta_{di}+\delta_{al}\delta_{bi}\delta
_{cj}\delta_{dk}\right)  \right]
\end{align}%
\end{subequations}%

\begin{figure}[tb] \centering
\includegraphics[width=40mm]{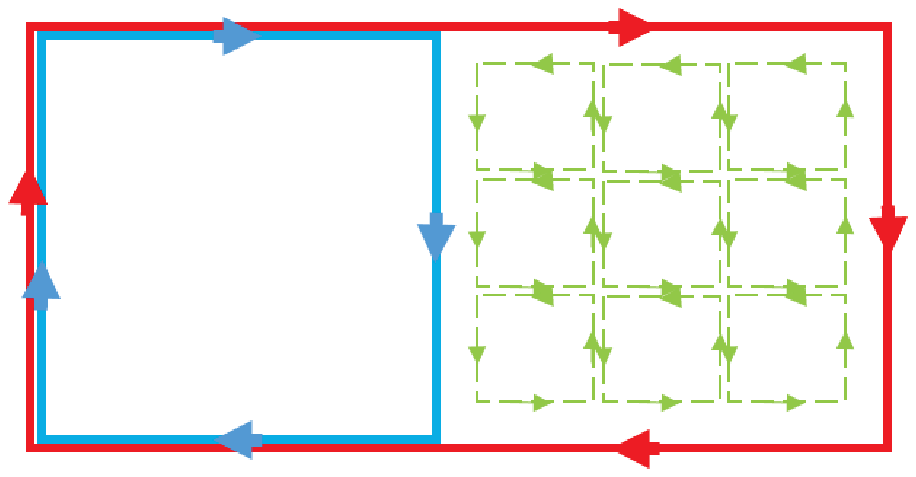}%
\includegraphics[width=50mm]{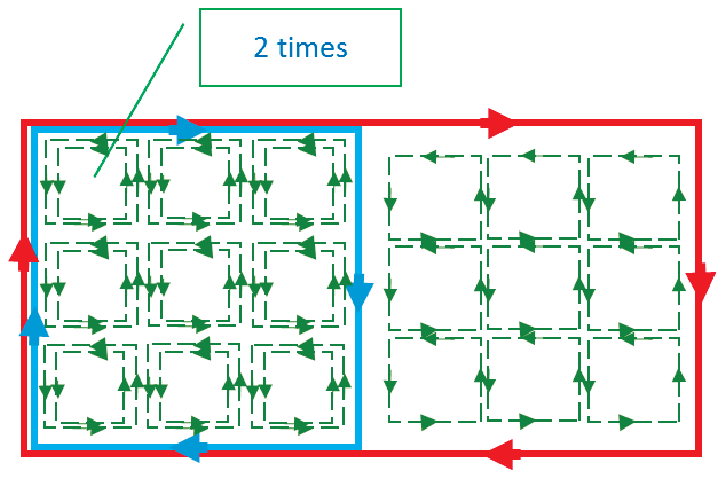}%
\caption{strong coupling expansion: contributing graph for $SU(2)$ case}%
\label{fig:SCexpansion}%
\end{figure}

Figure \ref{fig:SCexpansion} shows the examples which contribute to the
expectation of the double-winding Wilson loop. The leading term of the
expansion is given by the planner diagram that covers the minimal areas
$S_{1}$ and $S_{2}.$%

\begin{description}
\item[SU(2) case :] The leading contribution is given by the leftmost diagram
of Fig.\ref{fig:SCexpansion}. \ The area $S_{2}$\ is not fulfilled by
plaquettes, and we have only contribution from area $S_{1}-S_{2}$ . Thus we
have the difference-of-area behavior:
\begin{equation}
\left.  W(C)\right\vert _{\text{Leading}}=-2\left(  \frac{\beta}{N_{c}%
}\right)  ^{S_{1}-S_{2}}\text{.}\label{eq:SU2-leading}%
\end{equation}
The right diagram of Fig.\ref{fig:SCexpansion} is the higher order term, which
gives the sum-of-area behavior:
\begin{equation}
\left.  W(C)\right\vert _{\text{right}}=-q_{N_{c}}(S_{2})\left(  \frac{\beta
}{N_{c}}\right)  ^{S_{1}+S_{2}},\label{eq:SU2-C}%
\end{equation}
where the coefficient $q_{N_{c}}(S)$ is given by \
\begin{equation}
q_{Nc}(S)=\frac{N_{c}^{2}}{2}\left[  \left(  \frac{N_{c}}{N_{c}-1}\right)
^{S-1}-\left(  \frac{N_{c}}{N_{c}+1}\right)  ^{S-1}\right]  .\label{eq:Q}%
\end{equation}

As for $U(1)$ case it should be noticed that the average of the left panel of
Fig.\ref{fig:SCexpansion} vanishes in the $U(1)$ case, and the leading term
starts from the right panel. Therefore, we have the sum-of-area behavior.

\item[SU(3) case :] The leading term is given by the left panel of
Fig.\ref{fig:SCexpansio2}.%
\begin{equation}
\left.  W(C)\right\vert _{\text{Leading}}=-2\left(  \frac{\beta}{N_{c}%
}\right)  ^{S_{1}}\text{.}\label{eq:SU3-leading}%
\end{equation}
The higher corrections are given by the center and rightmost diagrams of
Fig.\ref{fig:SCexpansio2}
\begin{equation}
\left.  W(C)\right\vert _{\text{correction}}=-q_{N_{c}}(S_{2})\left(
\frac{\beta}{N_{c}}\right)  ^{S_{1}+S_{2}}-q_{N_{c}}(S_{1}-S_{2})\left(
\frac{\beta}{N_{c}}\right)  ^{2S_{1}-S_{2}}.\label{eq:SU3-C}%
\end{equation}
The area law is neither the neither difference-of-area behavior nor the
sum-of-area behavior.

\item[SU(4) case :] For $N_{c}=4,$ the left and center diagrams in
Fig.\ref{fig:SCexpansio2} give the same contribution:%
\begin{equation}
\left.  W(C)\right\vert _{\text{Leading}}=-2q_{N_{c}}(S_{2})\left(
\frac{\beta}{N_{c}}\right)  ^{S_{1}+S_{2}}.\label{eq:SC4}%
\end{equation}

\item[SU(N$_{c}$) $\dot{N}_{c}\geq5$ case :] For $N_{c}\geq5$, the leading
diagram is interchanged, and the center diagram in Fig.\ref{fig:SCexpansio2}
is the leading term:
\begin{equation}
\left.  W(C)\right\vert _{\text{Leading}}=-q_{N_{c}}(S_{2})\left(  \frac
{\beta}{N_{c}}\right)  ^{S_{1}+S_{2}}.\label{eq:SU(4<N)}%
\end{equation}%
\begin{figure}[tbp] \centering
\includegraphics[width=50mm]{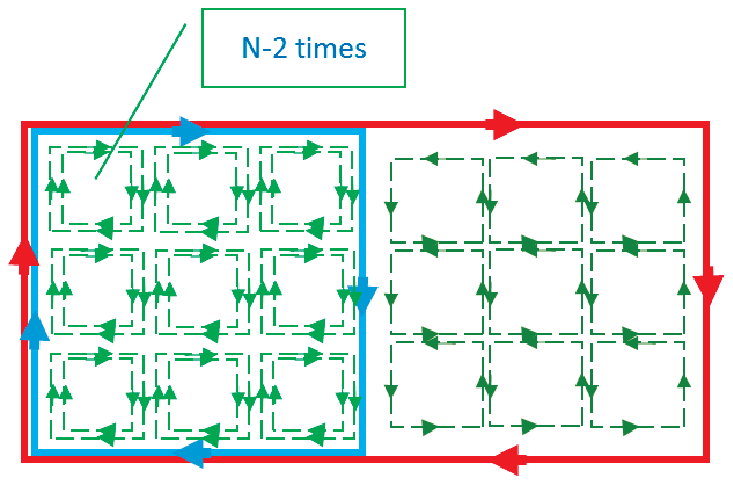}%
\includegraphics[width=50mm]{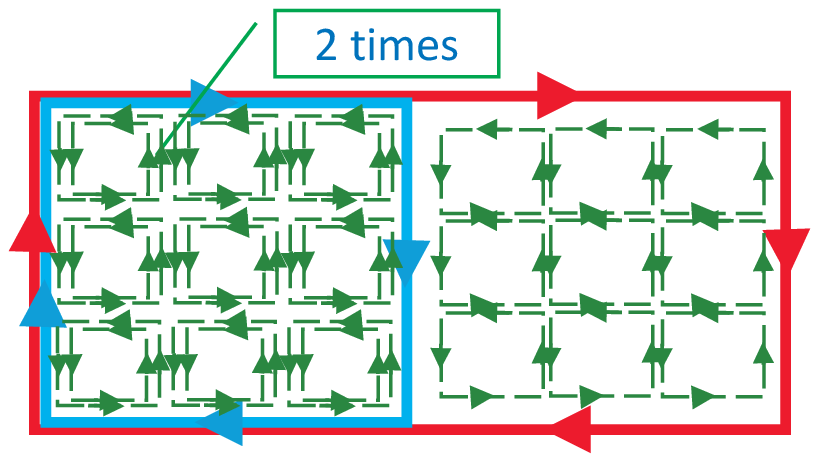}%
\includegraphics[width=50mm]{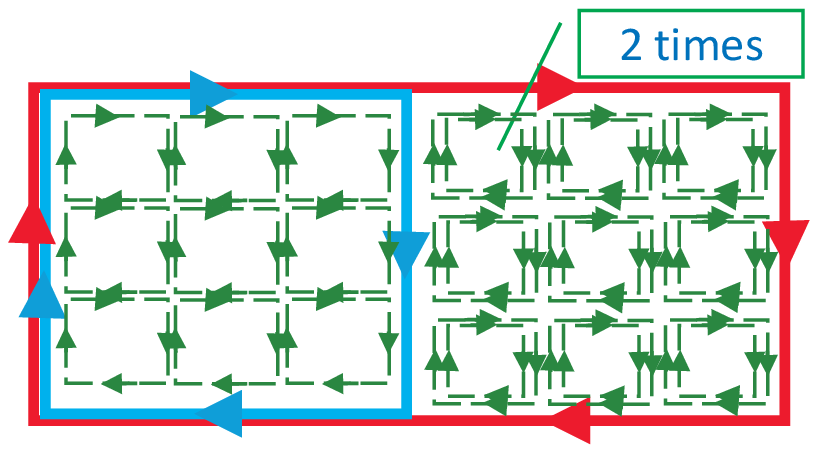}%
\caption{strong coupling expansion: contributing graph for
$SU(N)$ $N\ge3$ .  The rightmost panel is only for $SU(3)$ case }%
\label{fig:SCexpansio2}%
\end{figure}%

\end{description}

\section{Numerical simulation}

\label{sec:numarical} We perform the numerical simulation on the lattice by
using the Wilson action. For $SU(2)$ case, we generate 1000 configurations for
$32^{4}$ lattice with $\beta=2.6$ by using the standard pseudo heat-bath
method. For $SU(3)$ case, we generate 1000 configurations for $24^{4}$ lattice
with $\beta=6.2$ by using Cabibo-Marinari\cite{Cabibbo} and over-relaxation
algorithms. In the measurement of the Wilson loop average, the gauge links are
smeared by using the APE smearing method \cite{APE}.

\subsection{SU(2) case}

\label{sec:numerical-su2}%
\begin{figure}[tbp] \centering
\includegraphics[width=40mm, angle=270]{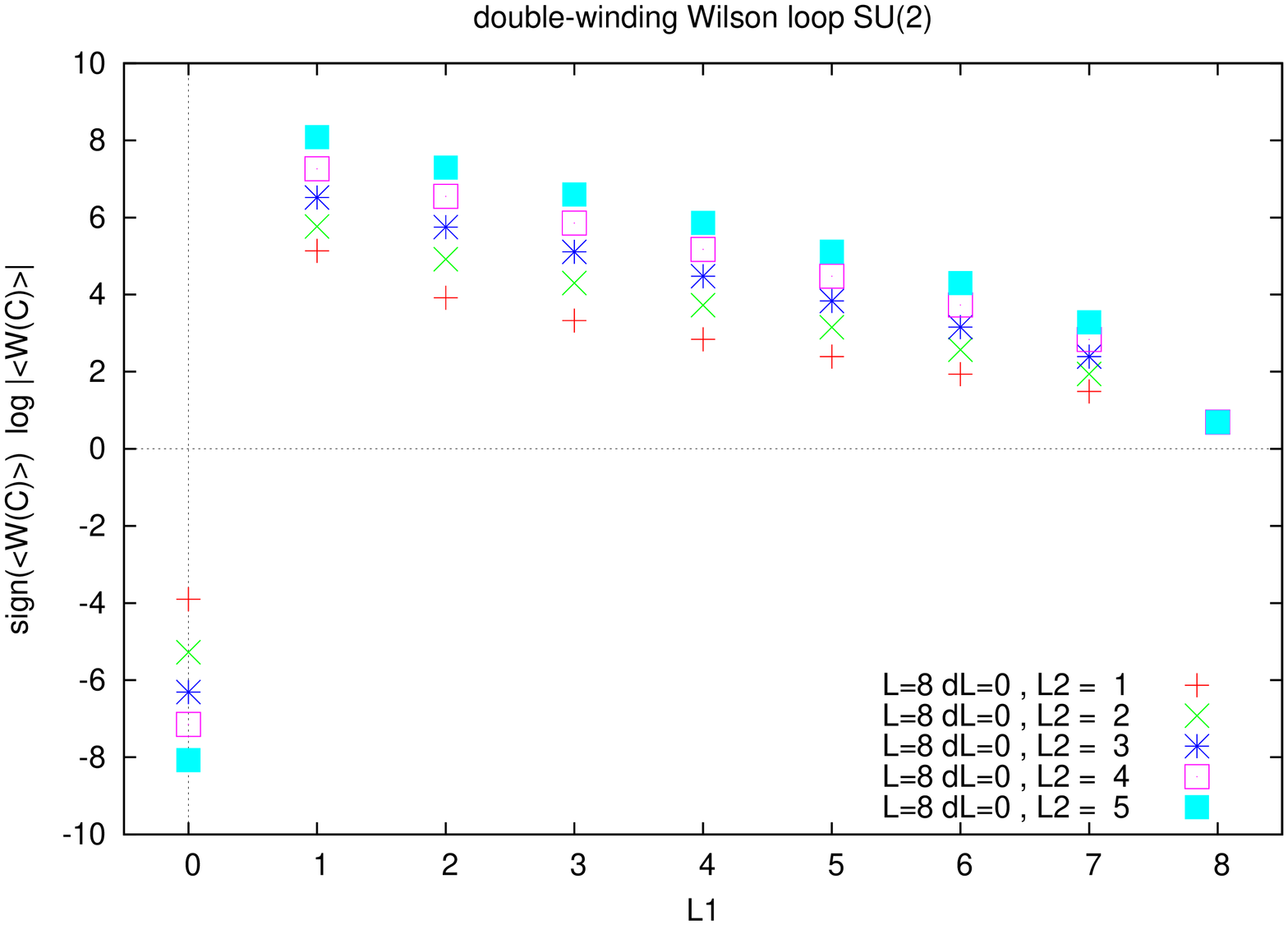}
\includegraphics[width=40mm, angle=270] {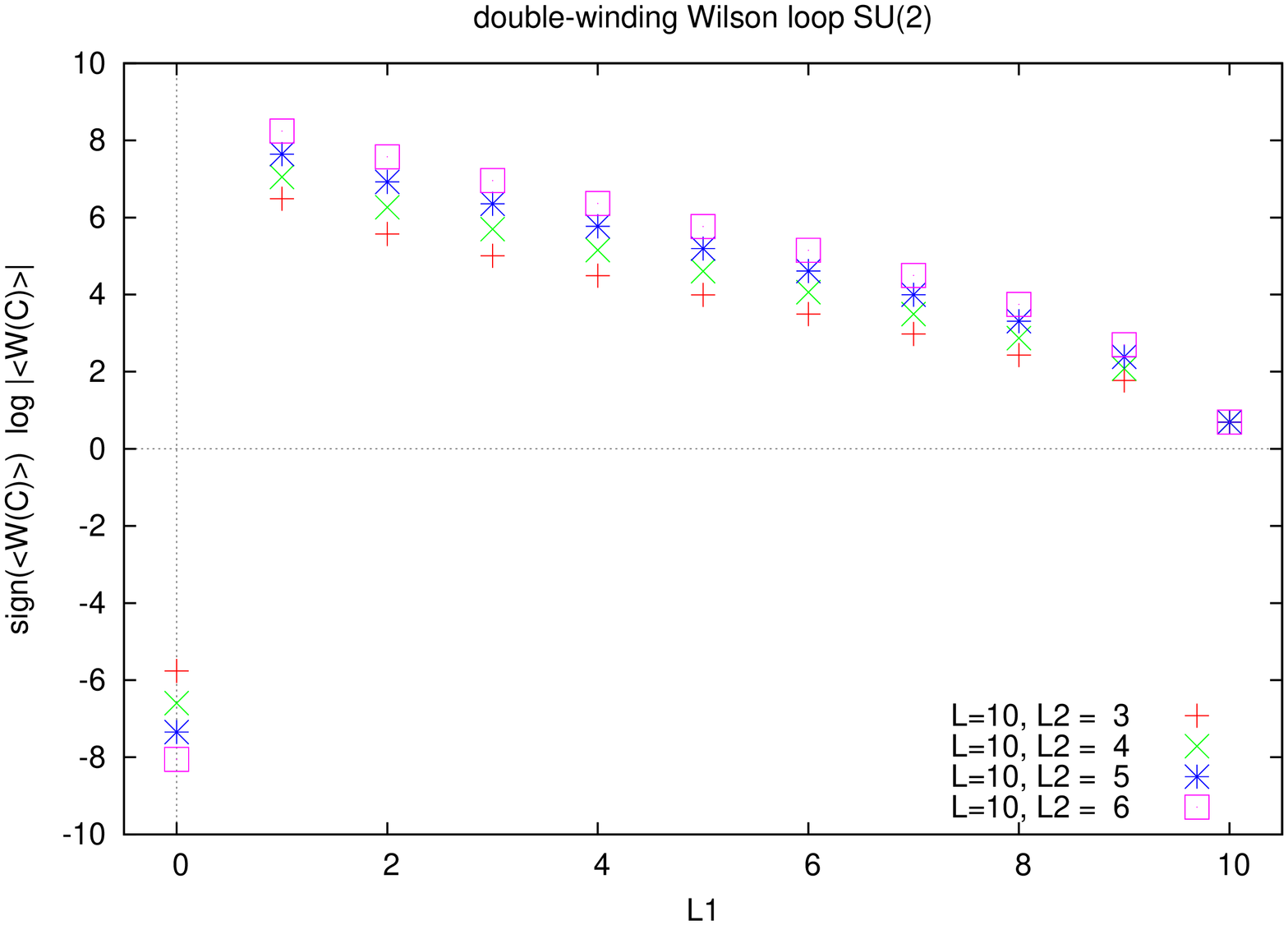}
\caption{
The measurement of the Wilson loop average of  Fig. \ref{fig:dwinding-wloop-L}
($\delta L=0$) for the $SU(2)$ case :
The left and right plots  represent  $ \text{sign} ( \left\langle
W(C) \right\rangle) \text{log} (| \left\langle W(C) \right\rangle
|) $  v.s. $L_1 $
for  $L=8$  and  $L=10$, respectively.  The case of $L_1=0$ corresponds to the single-winding Wilson loop ($C=C_1$),
and  the case of $L_1=L$ corresponds to  the double-winding Wilson loop with identical contours (the rightmost panel of
Fig. \ref{fig:dwinding-wloop-L} , $C_1=C_2$).  }
\label{fig:Sim-SU2}%
\end{figure}
First, we investigate the double-winging Wilson loop for the $SU(2)$ case. The
double-winding Wilson-loop operator, $W(C=C_{1}\times C_{2})$, is represented
at the center panel of Fig. \ref{fig:dwinding-wloop-L}. Note that the case of
$L_{1}=0$ corresponds to the single-winding Wilson loop ($C=C_{1}$), and the
$L_{1}=L$ case corresponds to the case of the two identical loops. We measure
the expectation value of the Wilson loop $\left\langle W(C)\right\rangle $ for
various $L_{1}$, $L_{2}$ with fixed $L$. As $L_{1}$ increases with fixed $L$
and $L_{2}$, $S_{1}$ is constant, $S_{2}$ increases, and then $S_{1}-S_{2}$
decreases. The result is shown in Figure \ref{fig:Sim-SU2}. The vertical axis
represents the logarithmic-scale Wilson loop average as $\mathrm{sign}%
(\left\langle W(C)\right\rangle )\log(\left\vert \left\langle
W(C)\right\rangle \right\vert )$. The left and right panels show the case of
$L=8$ and $L$ $=10,$ respectively.\ The Wilson loop average changes sign for
the case of the single-winding loop and the double-winding loop, i.e.,  in the
case of $L_{1}=0$,  the Wilson loop average $\left\langle W(C)\right\rangle $
takes positive value with $\left\vert \left\langle W(C)\right\rangle
\right\vert \leq1$, while, in the case of $L_{1}\neq0$, the Wilson loop
average\ takes a negative value with $\left\vert \left\langle
W(C)\right\rangle \right\vert \leq1$. The plots show that the absolute value
of the double-winding Wilson loop average falls off as $L_{1}$ increases. This
result is consistent with the result in the strong coupling expansion: The
Wilson loop average falls off as the difference-of-area behavior.

\subsection{SU(3) case}

\label{sec:numerical-su3}%
\begin{figure}[tbp] \centering
\includegraphics[width=40mm, angle=270]{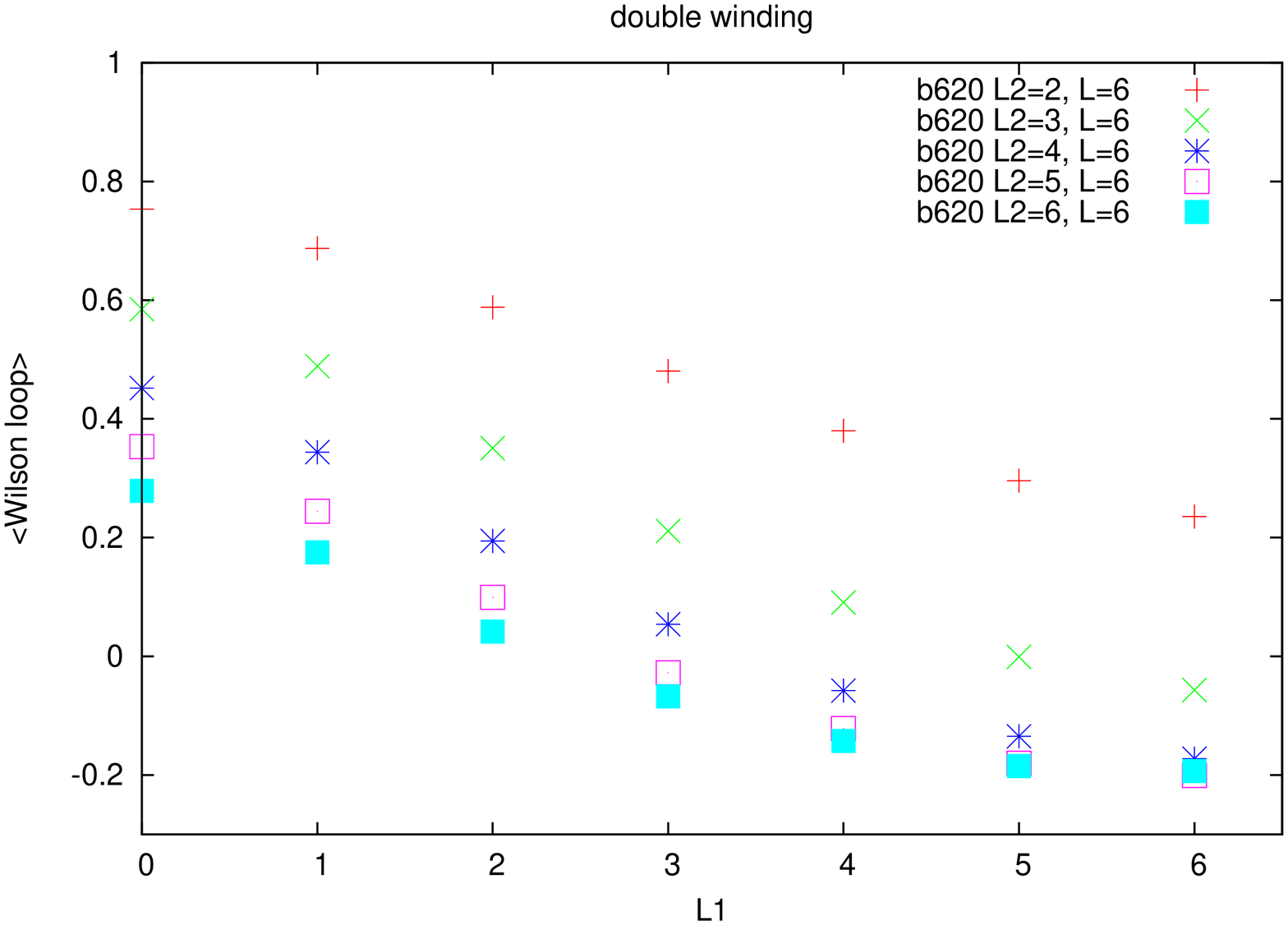}
\includegraphics[width=40mm, angle=270]{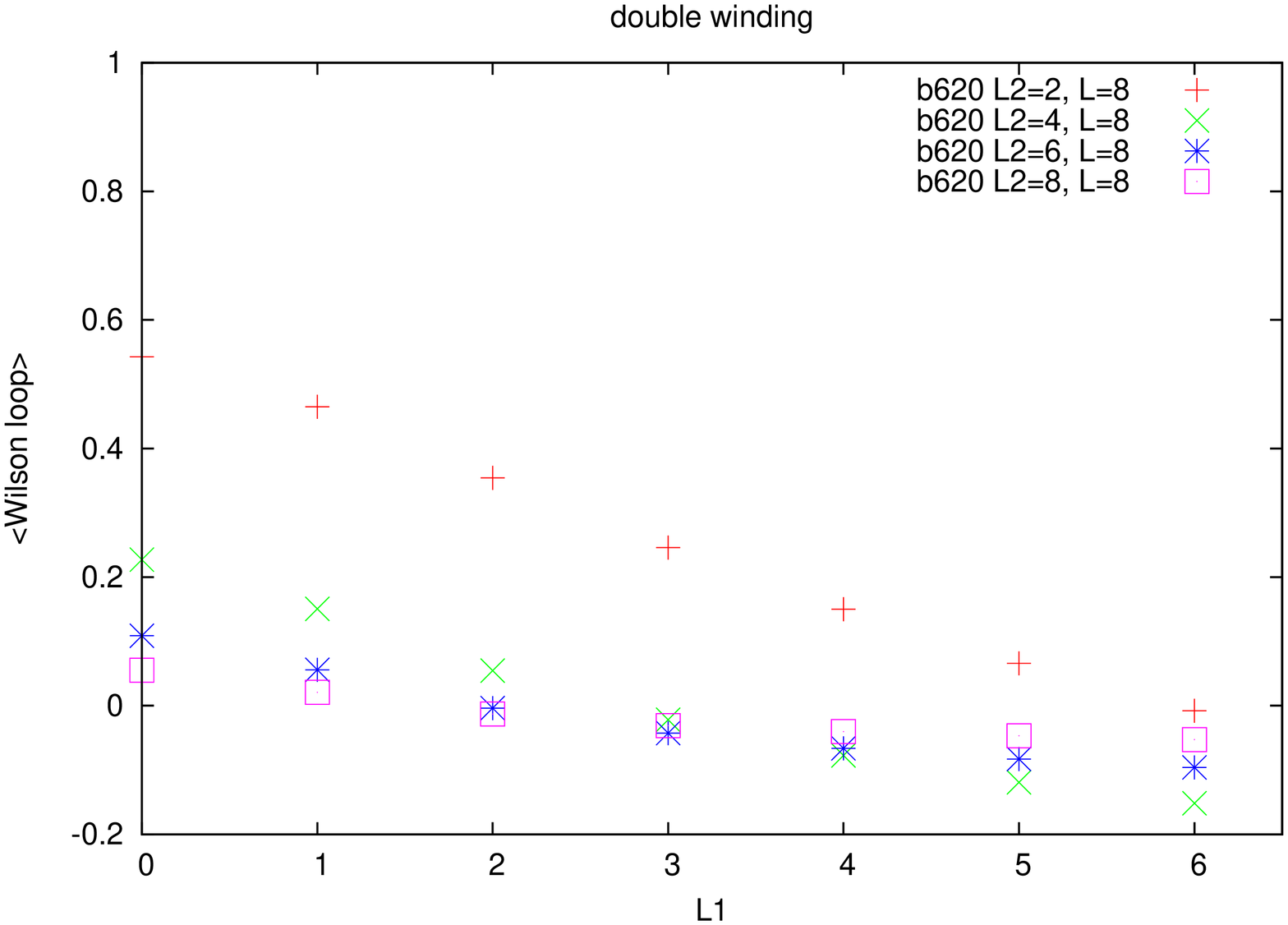}
\caption{
The measurement of the double-winding Wilson loop for the SU(3) case. ($\delta
L =0$)
The left and right panels show the plots $\left\langle W(C) \right
\rangle$ v.s. $L_1$ for
$L = 6 $ and  $L=8$, respectively.   The case of $L_1=0$ and $L_1=L$ represents
the single-winding Wilson loop, and  the double-winding Wilson loop with identical contours,
respectively.}%
\label{fig:sim-SU3}%
\end{figure}
We investigate the $SU(3)$ case. Figure \ref{fig:sim-SU3} shows the Wilson
loop average for $\delta L=0$, for various $L_{1}$ and $L_{2}.$ The left panel
shows the case of $\ L=6$. As $L_{1}$ increases, the Wilson loop average
decreases. For a small area of $S_{1}$ $\left\langle W(C)\right\rangle $ is
positive, while for large area of $\ S_{1}$ $\left\langle W(C)\right\rangle $
decreases to negative value as $L_{1}$ increases. The right panel shows the
case of $\ L=8$. As $L_{1}$ increases, the Wilson loop average decreases.
However, for large area of $S_{1}$, $\left\langle W(C)\right\rangle $ deceases
slowly from $\left\langle W(C_{1})\right\rangle $ ($>0$) to $\left\langle
W(C_{1}\times C_{1})\right\rangle $ ($<0$) as $L_{1}$ increases. In case of
large $S_{1}$ and $S_{2}$, the Wilson loop average $\left\langle
W(C)\right\rangle $ is negative and almost constant. The double-winding Wilson
loop for the $SU(3)$ case obeys the area law of $S_{1}.$This result is
consistent with the strong-coupling expansion.

\begin{figure}[tbp] \centering
\includegraphics[width=40mm, angle=270]{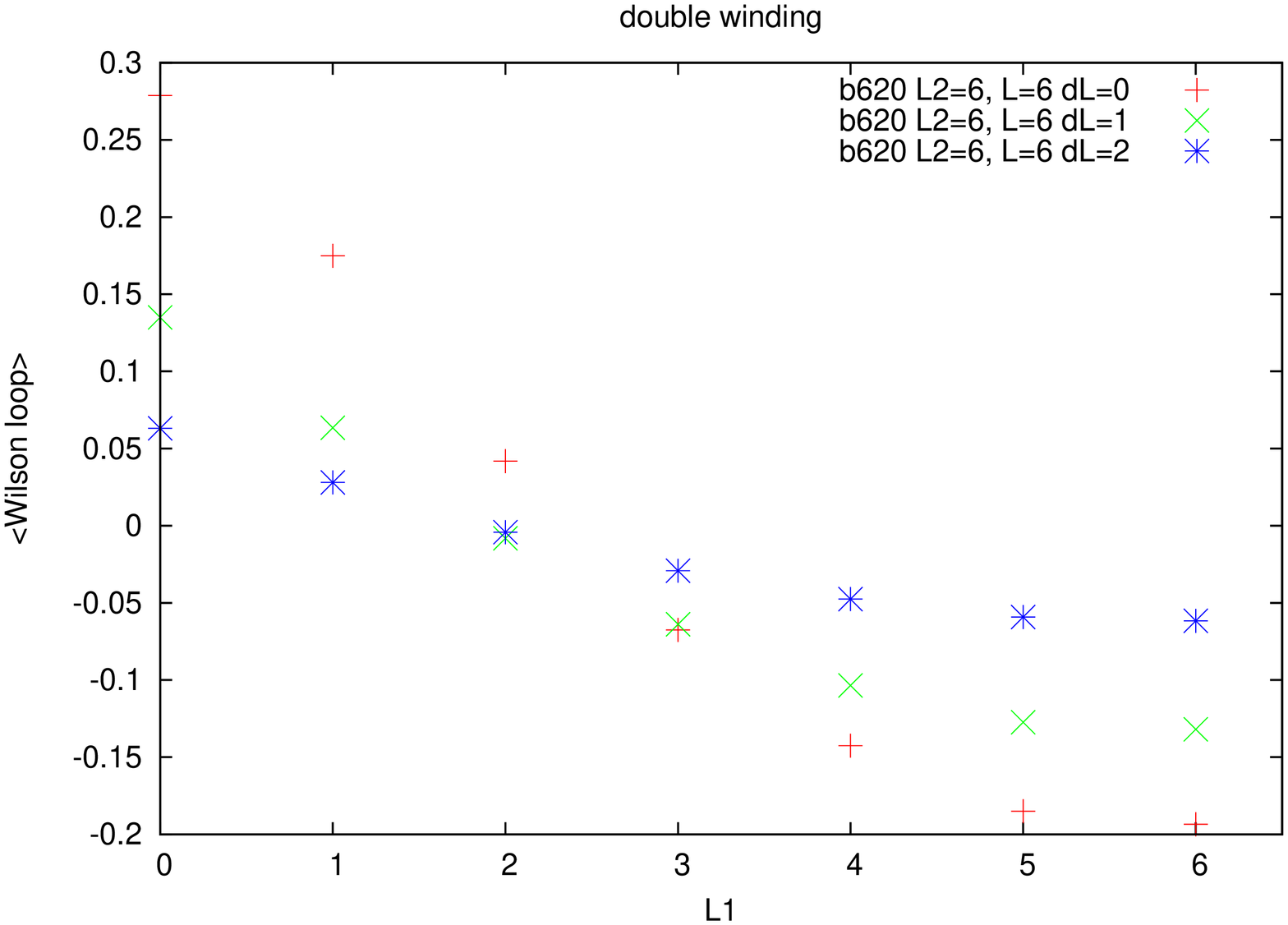}
\includegraphics[width=40mm, angle=270]{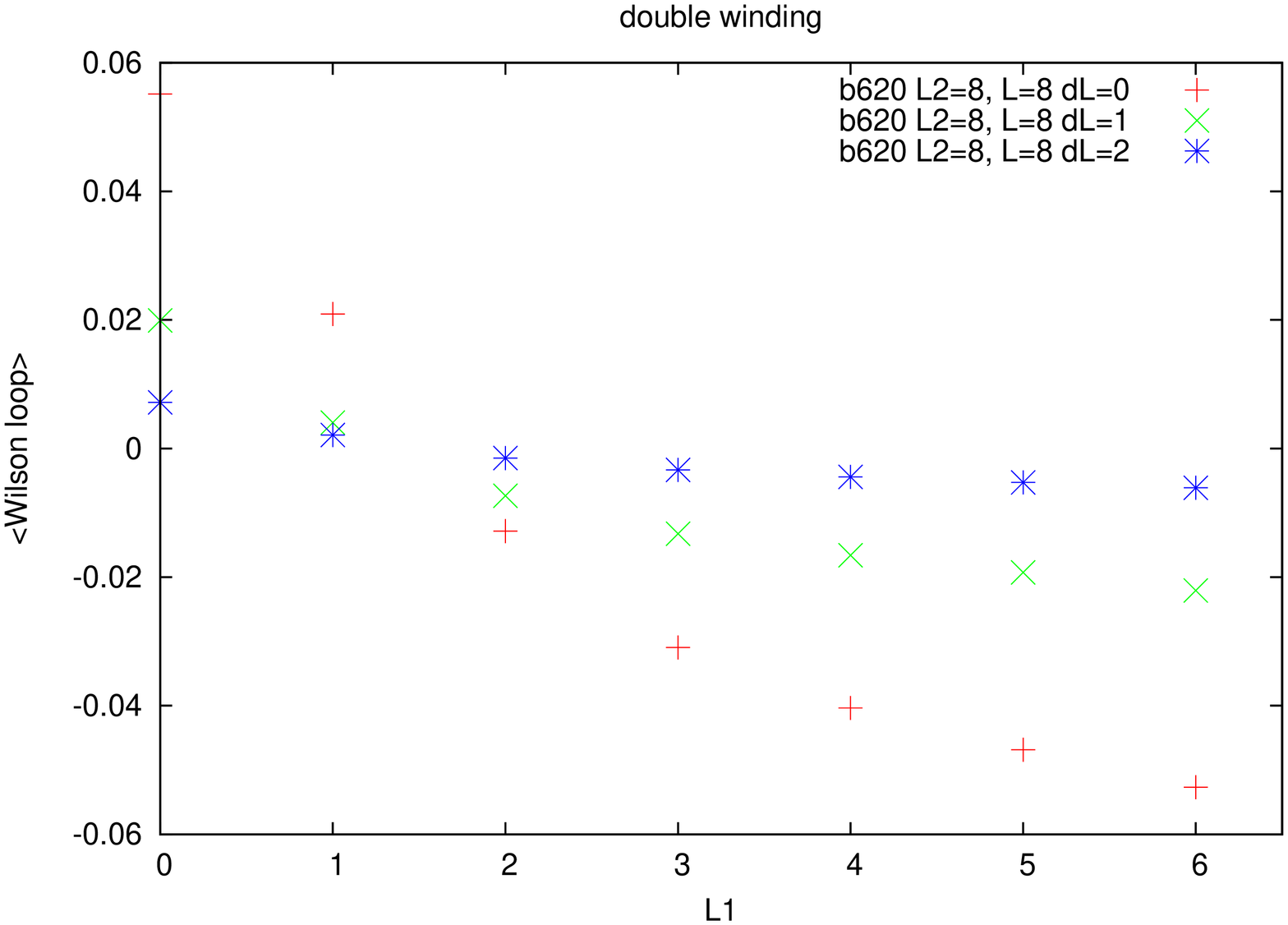}
\caption
{ The measurement of the double-winding Wilson loop average for the SU(3) case.
The left and right panel show  $\delta L$ dependence of
$\left\langle W(C) \right\rangle
$ v.s. $L_1$ for $L = L_2=6 $ and  $L=L_2=8$, respectively.
The case of $L_1=0$ corresponds to the single-winding Wilson loop. }%
\label{fig:sim-SU3-delta}%
\end{figure}

Next we investigate the case of $\delta L\neq0$ (see left panel of Fig.
\ref{fig:dwinding-wloop-L}). Figure \ref{fig:sim-SU3-delta} shows measurement
of the double-winding Wilson loop average for various $\delta L$. As is the
same with the case of $\delta L=0$, $\left\langle W(C)\right\rangle $ deceases
from $\left\langle W(C_{1})\right\rangle $ ($>0$) as $L_{1}$ increases. For
large $S_{1}$ and $S_{2},$ the Wilson loop average $\left\langle
W(C)\right\rangle $ is negative and almost constant. Therefore, the
double-winding Wilson loop average is independent \ of $L_{1}$ ($S_{2})$, and
it follows the area law of the area enclosed by $C_{1}$, i.e., $S_{1}$.%

\section{Double-winding Wilson loop with an identical contour}

We finally discuss the double-winding Wilson loop with an identical contour
($C_{1}=C_{2}=C$) (See the rightmost panel of Fig.\ref{fig:dwinding-wloop-L}).
This can be rewritten by using the Wilson loops in the irreducible
representations \cite{MastsudoKondo17} :%

\begin{subequations}%
\label{eq:WL}%

\begin{description}
\item[SU(2) case :] $2\otimes2=2\otimes2^{\ast}=1\oplus3$%
\begin{equation}
\left\langle W(C\times C)\right\rangle =-\frac{1}{2}+\frac{3}{2}\left\langle
W_{\text{adj}}(C)\right\rangle \label{eq:WLsu2}%
\end{equation}

\item[SU(3) case :] $3\otimes3=3^{\ast}\oplus6$%
\begin{equation}
\left\langle W(C\times C)\right\rangle =-\left\langle W_{[0,1]}%
(C)\right\rangle +2\left\langle W_{[2,0]}(C)\right\rangle \label{eq:WLsu3}%
\end{equation}

\item[SU(N) case :] $N\otimes N=\left(  \frac{N(N-1)}{2}\right)  _{A}%
\oplus\left(  \frac{N(N+1)}{2}\right)  _{S}$%
\begin{equation}
\left\langle W(C\times C)\right\rangle =-\frac{N-1}{2}\left\langle
W_{[0,1,....,0]}(C)\right\rangle +\frac{N+1}{2}\left\langle W_{[2,0,....,0]}%
(C)\right\rangle \label{eq:WLsuN}%
\end{equation}

\end{description}%

\end{subequations}
Here the representation is specified by the Dynkin indices, e.g., the
(anti)fundamental representation $\mathbf{3}^{\ast}$ with the Dynkin index
$[0,1]$, the sextet representation $\mathbf{6}$ with the Dynkin index $[2,0].$
If one assumes the Casimir scaling of the string tension, one can estimate the
double-winding Wilson loop average,
\begin{equation}
\left\langle W_{R}(C)\right\rangle \simeq\exp(-S\sigma_{R})\text{ \ \ with
}\sigma_{R}=\frac{C_{2}(R)}{C_{2}(F)}\sigma_{F},
\end{equation}
where $\sigma_{F}$, $\sigma_{R}$, $C_{2}(F)$, and $C_{2}(R)$ denote the string
tension of \ the fundamental representation ($F$) and the representation $R$,
and the quadratic Casimir operator of the fundamental representation and the
representation $R$, respectively.

In the calculation of the strong-coupling expansion, the Wilson loop average
can be estimated for large area $S$. Therefore, the Wilson-loop average in the
lower dimensional representation become dominant for large area $S$, that is,
the first term in each eq(\ref{eq:WL}) becomes dominant. These results are
consistent with results in strong coupling expansion.

\begin{figure}[tbp] \centering
\includegraphics[width=40mm, angle=270]{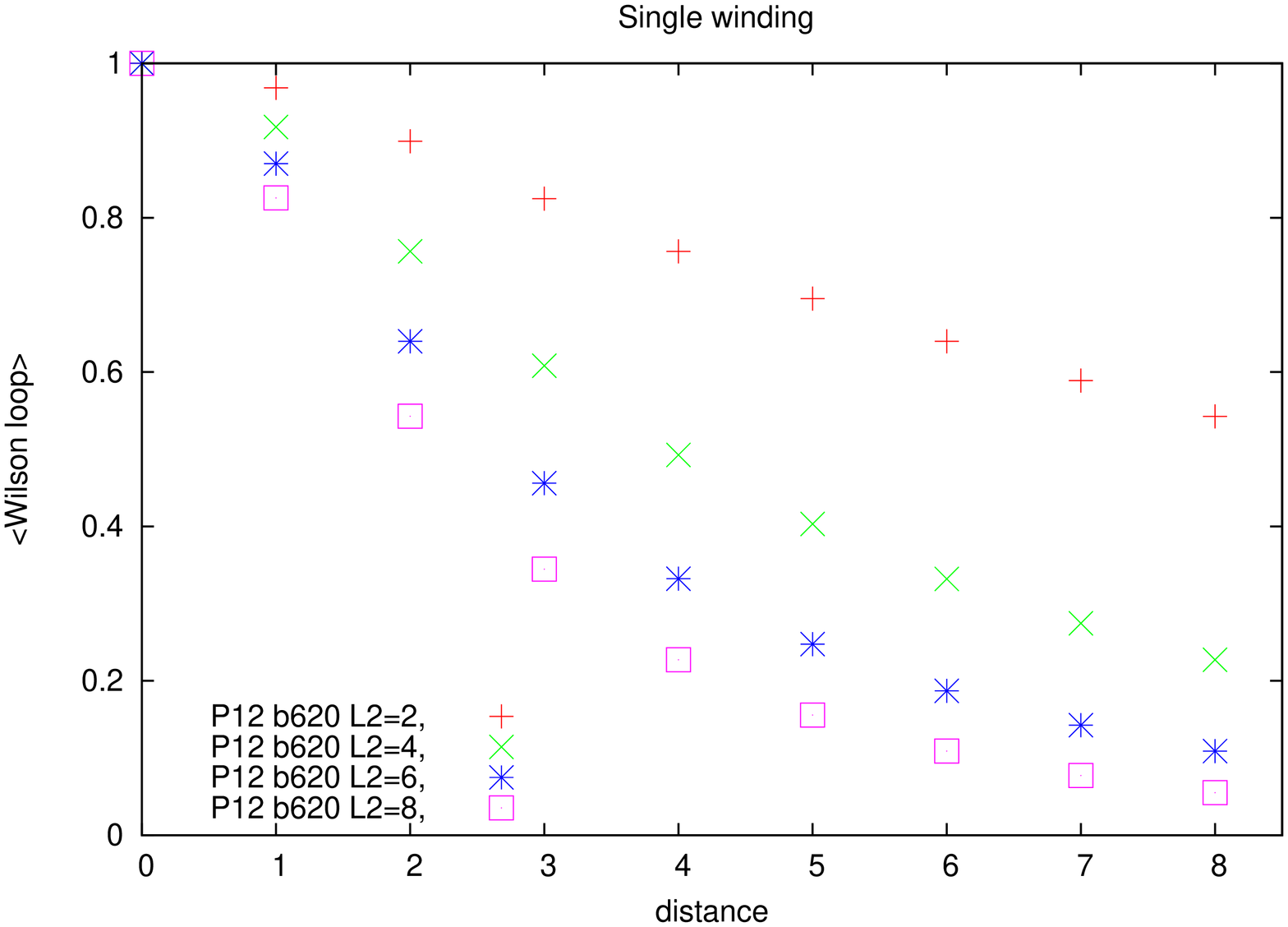}%
\includegraphics[width=40mm, angle=270]{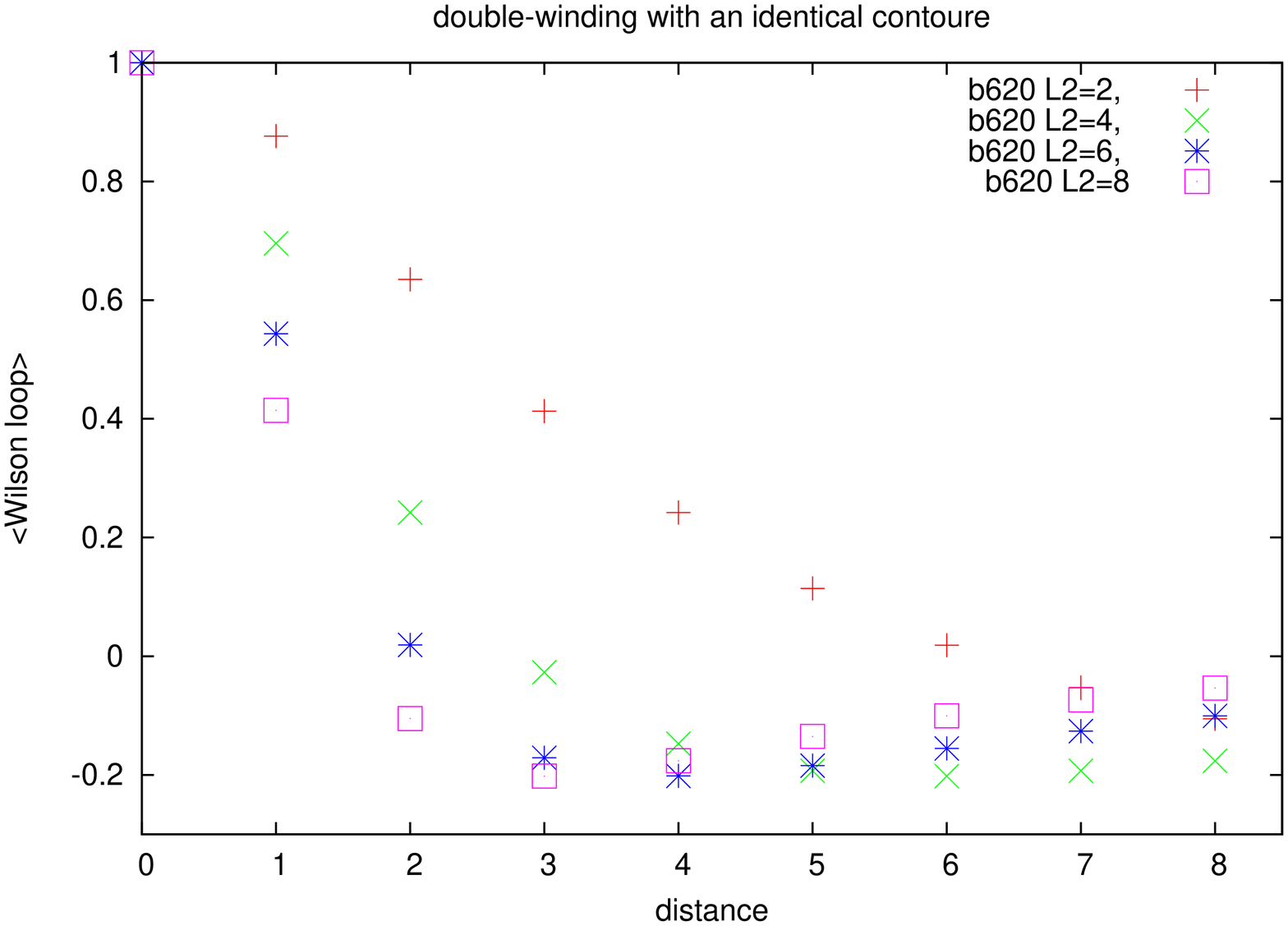}
\caption{
(left) The single-winding Wilson loop (right) the double-winding Wilson loop with identical contour. }%
\label{fig:sim-SU3-ident}%
\end{figure}

Next, we examine the relation eqs(\ref{eq:WL}) and the numerical simulations.
Figure \ref{fig:sim-SU3-ident} shows the Wilson loop averages of the
single-winding Wilson loop for the fundamental representation (left panel) and
\ the double-winding Wilson loop with identical contour (right panel) for the
$SU(3)$ case.\ The single-winding-Wilson-loop average of the representation
$R$ is positive and it falls off monotonically as the area $S$ increases.
However, the double-winding Wilson loop average decreases as the area $S$
increases, and changes the sign from positive to negative. As $S$ further
increases, the absolute value of the Wilson-loop average decreases to zero.
These are consistent with eq(\ref{eq:WLsu3}), because the second term
$\left\langle W_{[2,0]}(C)\right\rangle $ in eq(\ref{eq:WLsu3}), is dominant
for small $S$, and fall off quickly as $S$ increases. For larger $S$, the
dominant term is switched from the second one to the first one.

\section{Summary and discussion}

We have investigated the double-winding Wilson loop average for $SU(N_{c})$
Yang-Mills theory by using the strong coupling expansion and the lattice
simulation. By using the strong coupling expansion, we obtain the
difference-of-area \ behavior for the $SU(2)$ case.  For the $SU(N)$ ($N\geq
3$) case, however, the area law is the neither difference-of-area behavior nor
sum-of-area behavior. By using numerical simulation, we have confirmed the
result of the strong coupling expansion for the Wilson loop with large areas
$S_{1}$ and $S_{2}.$ These results are consistent with the results from the
continuum theory \cite{MastsudoKondo17} .

We are further interested in the dual superconductivity in the higher
dimensional representation of quarks. It has been pointed out that naively
replacing\ the Yang-Mills field with the Abelian projected field cannot
reproduces the correct result \cite{Greensite15}. In order to confirm the dual
superconductivity, we should go back to the non-Abelian Stokes theorem (NAST)
for the higher-dimensional representation \cite{MatsudoKondo15}, and derive
the Wilson-loop operator with the restricted ("Abelian") field that can
reproduce the area law in the NAST. These studies will appear in near future work.

\subsection*{Acknowledgement}

A.S and K.-I. K. were supported by Grant-in-Aid for Scientific Research, JSPS
KAKENHI Grant Number (C) No.15K05042. R. M. was supported by Grant-in-Aid for
JSPS Research Fellow Grant Number 17J04780. The numerical calculations are
supported by the Large Scale Simulation Program No.16/17-20(2016-2017) of High
Energy Accelerator Research Organization (KEK).

\end{document}